\documentclass [12pt] {report}
\usepackage{amssymb}
\newcommand {\eqdef} {\stackrel{\rm def}{=}}
\newcommand {\D}[2] {\displaystyle\frac{\partial{#1}}{\partial{#2}}}

\newcommand {\Dd}[3] {\displaystyle\frac{\partial^2{#1}}{\partial{#2}\partial{#3}}}
\newcommand {\al} {\alpha}

\newcommand {\ga} {\gamma}

\newcommand {\de} {\delta}
\newcommand {\De} {\Delta}

\newcommand {\prtl} {\partial}
\newcommand {\fr} {\displaystyle\frac}

\newcommand {\be} {\begin{equation}}
\newcommand {\ee} {\end{equation}}
\newcommand {\ba} {\begin{array}}
\newcommand {\ea} {\end{array}}
\newcommand {\bp} {\begin{picture}}
\newcommand {\ep} {\end{picture}}
\newcommand {\bc} {\begin{center}}
\newcommand {\ec} {\end{center}}
\newcommand {\bt} {\begin{tabular}}
\newcommand {\et} {\end{tabular}}
\newcommand {\lf} {\left}
\newcommand {\rg} {\right}

\newcommand {\cA} {{\cal A}}

\newcommand {\cC} {{\cal C}}

\newcommand {\cH} {{\cal H}}
\newcommand {\cI} {{\cal I}}

\newcommand {\cS} {{\cal S}}
\newcommand {\cT} {{\cal T}}

\newcommand {\cV} {{\cal V}}

\newcommand {\ses} {\medskip}

\newcommand {\bibit} {\bibitem}
\newcommand {\nin} {\noindent}
\newcommand {\set}[1] {\mathbb{#1}}

\def\2#1#2#3{{#1}_{#2}\hspace{0pt}^{#3}}
\def\3#1#2#3#4{{#1}_{#2}\hspace{0pt}^{#3}\hspace{0pt}_{#4}}

\newcounter{sctn}
\def\sec#1.#2\par{\setcounter{sctn}{#1}\setcounter{equation}{0}
                  \noindent{\bf\boldmath#1.#2}\bigskip\par}
\begin {document}

\begin {titlepage}

\vspace{0.1in}

\begin{center}
{\Large
Hypercomplex Numbers, Associated Metric Spaces,
}\\
\end{center}

\begin{center}
{\Large
and Extension of
Relativistic Hyperboloid
}\\
\end{center}

\vspace{0.3in}

\begin{center}

\vspace{.15in}
{\large D.G. Pavlov\\}
\vspace{.25in}
{\it Division of Applied Mathematics, Academy of
Civil Defence.\\
Moscow, Russia\\

pavlovdg@newmail.ru

}

\vspace{.05in}

\end{center}

\begin{abstract}

We undertake
to develop  a
successful framework for commutative-associative
hypercomplex numbers
with the view to explicate and
study associated
 geometric and generalized-relativistic concepts,
basing on an
interesting possibility
to introduce appropriate
multilinear metric forms in the treatment.
The scalar polyproduct,
which extends the ordinary
scalar product used in bilinear (Euclidean
and pseudo-Euclidean) theories,
has been proposed and applied to be a generalized
metric base for the approach.
A fundamental concept of
multilinear
isometry
is
proposed.
This renders possible to muse upon
various relativistic
physical applications based on anisotropic {\it versus}
ordinary spatially-rotational case.

\end{abstract}

\end{titlepage}

\vskip 1cm

{\nin\bf 1. INTRODUCTION}
\medskip

William Hamilton's discovery of quaternions in 1843
was the first event of history when  the concept of
complex numbers has been successfully generalized.
Although the quaternions, as being  comprised of the ordered
quadruples
of real numbers possessing a convenient
noncommutative multiplication law, were playing
a
historically
important part in development of
both the algebra and the geometry,
they were the methods of
vector analysis which really streamlined powerful applications.
As a farther extension of quaternions, the so-called octaves were
developed after Cayley, - a striking particular feature was
occured
to be
an absence
of
 any associativity in the involved multiplications.
Recently,
many articles in the hypercomplex literature were focused on the adoption of
quaternionic methods to Dirac equation and to quantum mechanics
(see
the works
[1-23]
and references therein).
The polynumbers to be dealt with,
 like
ordinary
complex or bicomplex
numbers,
share many important properties of ordinary real numbers,
including
commutability, associativity, distributivity, existence of
zero and unity;
but also has important differences, namely,
presence of divisors of zero and absence of ordering.

It proves possible to
define the concept of a polyform to serve as a norm
for members of the
polynumber algebra,
which in turn makes us develop
a geometry for such
the algebra
 by assigning
lengths to associated
vectors
by means of  values of the respective
radicals
of the polyform.
Remarkably,
the geometric structure of the space can be extended even
farther
on the basis of
the
symmetrical multilinear form,
a scalar polyproduct in a sense,
which can be proposed to generalize the polyform.
Implying possible relativistic physical applications,
we shall restrict our treatment to the dimension $N=4$,
so that the
quadrahyperbolic numbers will be
the subject.

The new facet that has been opened in this vein is the metric
relationship
with the Finsler-Geometry presupposition.
Namely,
when we assign the length to a polynumber-associated vector by means
of the 4-th radical of the 4-th degree
polyform, we get the length which obeys the property of homogeneity of
degree one.
The latter property, however,
is a corner-stone for the Finsler Geometry
and for its modern relativistic applications [24-29].
Therefore,
there can be arranged a one-to-one
correspondence between
the set of
quadrahyperbolic numbers
and
the vector
space
 of a Finsler
type.
In this way, the equation of an
Finslerian
indicatrix
is tantamount
to the expression for the modulus of a unit
quadrahyperbolic number.
In a sense,
the respective
Finsler
space
acquires a
qualitatively new
property: its vectors can be not only summed
or deducted, but also multiplied,
and even divided (though the last operation
is unambiguous only if the length of a divisor differs
from
zero).

If it is possible to indicate a basis such that the squares of
the basis elements
become equal  $+1$, $0$, or $-1$, then one can conventionally says that
the algebra is of a {\it square-type} nature.
Such polynumbers may be of
hyperbolic, elliptic,
parabolic,
or mixed
type.
In the hyperbolic case the
squares
of all basis units are equal to $+1$, in the elliptic case
there is at least one
unity which square is
$-1$, and in the
parabolic case
there is at least one
member for
which
the square is zero. For mixed
polynumbers, basis units of each type
may occur.
The ordinary
real,
complex, double,
and dual numbers are particular cases of square-type
polynumbers.
It is natural
to expect that each interesting class of polynumbers possesses
a powerful geometric representation in an $n$-dimensional
vector
space.

The comparison between the ordinary hyperboloid
and
our
$\cI_4$-surface
clearly watersheds between the
pseudo-Euclidean traditional framework and the present
$\cH_4$-approach:
the anisotropic space-time is set forth $\cV\cS$
the spatially-isotropic idealization.
Actually, the
$\cH_4$-approach
starts with stipulation that the total space-time manifold exhibits four
distinguished directions, - which can be matched to the fact that
the real physical world exhibit many anisotropies as judged on
local as well on
cosmic scales.
For distribution of matter
and
energy
in the Universe is far from being isotropic.
Then
the
$\cH_4$-approach
may serve the purpose to reflect
geometrically and
algebraically
the anisotropic circumstances.
In this vein
the set $1,I,J,K$
acquires the
algebraic interpretation to mean
the hyperunits which
corresponds to
the anisotropic directions.
\ses

\nin
The work has been structured as follows.

We shall
deal mainly
with the four-dimensional case, which is of the particular importance
for physical relativistic applications.
In Section 2,
we introduce initial definitions
for
respective
quadrahyperbolic numbers,
 emphasizing the existence of attractive
definitions
for the mutual conjugates,
the polyform, and the norm.
The introduced concept of
mutual conjugacy, though being similar to that
applicable
in  case of complex, double,
and dual numbers, does not correspond to the concept of conjugacy
used for quaternions and octaves,
-
for the latters are not polynumbers in our sense because
their multiplication is noncommutative.
Many aspects and representations get simplified significantly on using
the absolute basis which is formed by the divisors of zero.
Section 2 is finished with the exponent representation for the
quadrahyperbolic numbers
under study,
which involves for them
the exponential arguments.
In Section 3, the consideration is transgressed into the isomorphic vector
space, whence the implications of associated metric forms
are of our particular
attention. Using the forms, we propose in Section 4
the fundamental concepts of the transversality, the orthogonality, and
the angle
for
the
quadrahyperbolic numbers
under study.
In particular, a due generalization of the Pythagorean theorem
is derived.
In Section 5,
we propose necessary definitions
to generalize concepts of arcs, triangles, and cones,
and also introduce the fundamental notion of isometry.

Is it possible to alternate the theory of ordinary
complex holomorphic functions, developed on the basis of the Cauchy-Riemann
equations, to get a successful holomorphic
hypercomplex-valued
functions
adapted
to
the
quadrahyperbolic numbers?
A due positive answer to the question can be found in Section 6.
The
Finslerian
topics covered briefly in Section 7.
In the last Section 8, basic points of our approach and relevant out-looks
will be emphasized.
In the appended section Research Problems we shall list some
directions in which one can see subject develop.
\ses
\ses\\

\setcounter{sctn}{2}
\setcounter{equation}{0}

{\nin\bf 2. Basic Definitions}
\ses

We introduce
{\it quadrahyperbolic numbers}
\be
A=
a_1'1
+
a_2'I
+
 a_3'J
+
 a_4'K,
\ee
where
$a_p'$
are real numbers called
{\it the components}, and $1,I,J,K$
are
{\it
basis units}.
Given another member
\be
B=
 b_1'1
+
 b_2'I
+
 b_3'J
+
 b_4'K,
\ee
the sum and the product of the two such numbers
are respectively defined as
\be
C=A+B
\ee
with
\be
C=(a_1'+b_1')1+(a_2'+b_2')I
+(a_3'+b_3')J+(a_4'+b_4')K,
\ee
and
as
\be
D=AB
\ee
with
$$
D=(a_1'b_1'+a_2'b_2'+
a_3'b_3'+
a_4'b_4')1+
(a_1'b_2'+a_2'b_1'+
a_3'b_4'+a_4'b_3')I
$$
\ses
\be
+(a_1'b_3'+a_2'b_4'+a_3'b_1'+a_4'b_2')J+
(a_1'b_4'+a_2'b_3+a_3'b_2'+a_4'b_1')K,
\ee
whenever the multiplication
table of basis units is
\begin{equation}
\ba{c|cccc}
&1&I&J&K\\
\hline
1&1&I&J&K\\
I&I&1&K&J\\
J&J&K&1&I\\
K&K&J&I&1\\
\ea
\end{equation}
It follows from this table that
\be
I^2=J^2=K^2=1.
\ee
Whence we  obtain the algebra
of commutative and associative
hypercomplex
numbers,
in which the
 multiplication
table
allows
to
qualify the corresponding
hypercomplex numbers as {\it hyperbolic},
to be referred to as
$\cH_4$-numbers.
Under these conditions, we say that we deal with
{\it the quadrahyperbolic algebra}, to be denoted as
$\cA\cH_4$.

We shall proceed by the following important
\ses

\nin
{\bf DEFINITION}. Members of a set $A_1,A_2,A_3,A_4$
are called
{\it
mutually
conjugate},
if the symmetric polynomials
\be
 A_1+A_2+A_3+A_4=P_1,
\ee
\ses
\be
 A_1A_2+A_1A_3+A_1A_4+A_2A_3+A_2A_4+A_3A_4=P_2,
\ee
\ses
\be
 A_1A_2A_3+A_1A_2A_4+A_1A_3A_4+A_2A_3A_4=P_3,
\ee
and
\be
 A_1A_2A_3A_4=P_4
\ee
are real.

We also introduce
\ses

\nin
{\bf DEFINITION}.  The product $P_4$ given by (2.12)
is called {\it the polyform}
of the number $A\in\cA\cH_4$, with
 $A\equiv A_1$.

Next, it is possible
to define the concept of {\it the quadrahyperbolic number's modulus}
\be
|A|=\sqrt[4]{|A_1A_2A_3A_4|}\,.
\ee
This satisfies the ordinary
properties
of a modulus:
\be
|\lambda A|=|\lambda||A|
\ee
and
\be
|AB|=|A||B|,
\ee
where $\lambda$ is a real number; $A$ and $B$ are two
hypercomplex numbers.

Given a number $A\in \cA\cH_4$,
it is
easy
and attractive
 to propose the following explicit set of mutual conjugates:
\be
A_1=a_1'+a_2'I+a_3'J+a_4'K,
\ee
\ses
\be
A_2=a_1'-a_2'I+a_3'J-a_4'K,
\ee
\ses
\be
A_3=a_1'+a_2'I-a_3'J-a_4'K,
\ee
\ses
\be
A_4=a_1'-a_2'I-a_3'J+a_4'K,
\ee
where $A_1\equiv A$.
The direct calculations yield
\be
A_1+A_2+A_3+A_4=4a_1',
\ee
\ses
\ses
\be
A_1A_2+A_1A_3+A_1A_4+A_2A_3+A_2A_4+A_3A_4=
6{a_1'}^2-2{a_2'}^2-2{a_3'}^2-2{a_4'}^2,
\ee
\ses
\ses
\be
A_1A_2A_3+A_1A_2A_4+A_1A_3A_4+A_2A_3A_4=4{a_1'}^3-4a_1'{a_2'}^2-4a_1'{a_3'}^2
-4a_1{a_4'}^2+8a_2'a_3'a_4',
\ee
\ses
\ses
$$
A_1A_2A_3A_4={a_1'}^4+{a_2'}^4+{a_3'}^4+{a_4'}^4
-2{a_1'}^2{a_2'}^2
-2{a_1'}^2{a_3'}^2
-2{a_1'}^2{a_4'}^2
$$
\ses
\ses
\be
-2{a_2'}^2{a_3'}^2
-2{a_2'}^2{a_4'}^2
-2{a_3'}^2{a_4'}^2
+8a_1'a_2'a_3'a_4'.
\ee

The fact that the product of mutual conjugates is a real number
implies the
possibility
to define {\it the operation of division}, treated as an operation inverse
to multiplication. So the inverse $A^{-1}$ of a given number $A$ is the
number
\be
A^{-1}\eqdef\frac{A_2A_3A_4}{P_4}.
\ee
Inverses exist only for those numbers whose polyform $P_4$,
as given by (2.12),
differs from zero.
If a number is not zero, but it's
polyform value is equal to zero,
then the number is called
a {\it divisor of zero}.

In the algebra $\cA\cH_4$,
we can propose
the following four explicit
divisors of zero
\be
S_1=\frac{1}{4}(1+I+J+K),
\ee
\ses
\be
S_2=\frac{1}{4}(1-I+J-K),
\ee
\ses
\be
S_3=\frac{1}{4}(1+I-J-K),
\ee
\ses
\be
S_4=\frac{1}{4}(1-I-J+K)
\ee
with the simple algebraic properties:
$S_iS_j=S_i,$ if $i=j,$ and $S_iS_j=0,$ if $i\ne j.$
We will call such
divisors
{\it absolute},
and the basis
consisting of them will be called an
{\it absolute basis.}

Inversely, the units $1,I,J,K$  can readily be
expressed through the divisors
\be
1=S_1+S_2+S_3+S_4,
\ee
\ses
\be
I=S_1-S_2+S_3-S_4,
\ee
\ses
\be
J=S_1+S_2-S_3-S_4,
\ee
\ses
\be
K=S_1-S_2-S_3+S_4.
\ee

Numbers from $\cA\cH_4$, when
written with respect to the absolute basis, can be easily
multiplied and divided.
Indeed, the product of two numbers $A$ and $B$ will be
\be
AB=(a_1b_1)S_1+(a_2b_2)S_2+(a_3b_3)S_3+(a_4b_4)S_4
\ee
and their quotient will read
\be
\frac{A}{B}=
\frac{a_1}{b_1}S_1+\frac{a_2}{b_2}S_2+\frac{a_3}{b_3}S_3+\frac{a_4}{b_4}S_4.
\ee
Usage of the absolute basis reveals the remarkable fact that
 the structure of the algebra $\cA\cH_4$
is isomorphic to the algebra of real diagonal
matrices,
for
 the set of mutual conjugates
(see
(2.9)-(2.12))
takes on the form
\be
A_1=a_1S_1+a_2S_2+a_3S_3+a_4S_4,
\ee
\ses
\be
A_2=a_2S_1+a_1S_2+a_4S_3+a_3S_4,
\ee
\ses
\be
A_3=a_3S_1+a_4S_2+a_1S_3+a_2S_4,
\ee
\ses
\be
A_4=a_4S_1+a_3S_2+a_2S_3+a_1S_4.
\ee
Therefore, the polyform of a number in the absolute basis
reads merely as
\be
P_4=a_1a_2a_3a_4,
\ee
and
 the simple expression
\be
|A|=\sqrt[4]{|a_1a_2a_3a_4|}\,
\ee
is obtained for the modulus of a quadrahyperbolic number.

Various
functions can be
introduced
on the set of polynumbers
$A\in\cA\cH_4$
by the help of series.
First of all,
the exponential function can be
be
defined
as
\be
e^X=1+X+\frac{X^2}{2}+\dots,
\ee
where $X$ is an arbitrary polynumber.
The
associated logarithmic function can be
given quite
traditionally:
\be
\ln X={\rm inverse\, of}\,(e^X).
\ee
Also,
\be
\cosh X=1+\frac{X^2}{2!}+\dots,\qquad
 \sinh X=X+\frac{X^3}{3!}+\dots\,.
\ee

Any number $A=a_1S_1+a_2S_2+a_3S_3+a_4S_4$,
whenever
the
components
$a_p$ are all greater than zero
with respect to the absolute basis,
can be represented by
 the following {\it exponential form}:
\be
A=|A|e^{\alpha I+\beta J+\gamma K},
\ee
\ses
\be
|A|=e^{\de},
\ee
where $|A|$ is the modulus (2.40);
the
real numbers
$\alpha,\beta$,
and $\gamma$,  similarally as in case of complex and double
numbers, may be called {\it the
arguments} of a
quadrahyperbolic number $A$;
the real number $\de$ represents the modulus.
This entails
\be
\alpha=\frac{1}{4}\ln\frac{a_1a_3}{a_2a_4}=
\frac{1}{4}(\ln a_1-\ln a_2+\ln a_3-\ln a_4),
\ee
\ses
\be
\beta=\frac{1}{4}\ln
\frac{a_1a_2}{a_3a_4}=
\frac{1}{4}
(\ln a_1+\ln a_2-\ln a_3-\ln a_4),
\ee
\ses
\be
\gamma=\frac{1}{4}\ln\frac{a_1a_4}{a_2a_3}=\frac{1}{4}(\ln a_1-\ln a_2-\ln a_3+\ln a_4),
\ee
\ses
\be
\de
=
\fr14\ln(a^1a^2a^3a^4)
=\fr14(\ln a^1+\ln a^2+\ln a^3+\ln a^4),
\ee
where $\ln x$ is the ordinary
logarithmic function of a real argument $x$.
\ses
Inversely,
\be
\ln a^1=\de+\al+\beta+\ga,
\ee
\ses
\be
\ln a^2=\de-\al+\beta-\ga,
\ee
\ses
\be
\ln a^3=\de+\al-\beta-\ga,
\ee
\ses
\be
\ln a^4=\de-\al-\beta+\ga.
\ee

Since each imaginary unit satisfies the hyperbolic analog of Euler's
formula:
\be
e^{\alpha I}=\cosh\alpha+I\sinh\alpha,
\ee
the exponent of an arbitrary quadrahyperbolic number
$X=\delta+\alpha I+\beta J+\gamma K$ can be given as
\be
e^X=(\cosh \delta+\sinh\delta)(\cosh\alpha+I\sinh\alpha)
(\cosh\beta+J\sinh\beta)(\cosh\gamma+K\sinh\gamma),
\ee
where $\cosh x$  and $\sinh x$ are ordinary
hyperbolic cosine and sine,
respectively.
\ses
\ses\\

\setcounter{sctn}{3}
\setcounter{equation}{0}

{\nin\bf 3. Associated Quadrahyperbolic Vector Space ${\bf\cV\cH_4}$}
\ses

The algebra
$\cA\cH_4$
can be juxtaposed by
{\it
the vector space
$\cV\cH_4$
}
in a
quite obvious way.
Having assumed a fixed basis in
$\cV\cH_4$,
we
shall denote respective images of
$\cH_4$-numbers
by bolds:
\be
{\bf A}\sim A,\quad
{\bf A}\in
\cV\cH_4,
\quad
A\in
\cA\cH_4.
\ee
In this vein, the norm should be written as
\be
||{\bf A}||=({\bf A,A,A,A}),
\ee
the fourth degree radical
\be
|\bf {A}|=\sqrt[4]{|({\bf A,A,A,A})|}\
\ee
will play the role of the length
of the vector ${\bf A}$,
and
the corresponding
symmetrical
{\it quadralinear form}
$({\bf A,B,C,D})$
can easily
be introduced
such that,
with respect to
{\it the absolute basis}
\be
{\bf S_1} = (1,0,0,0),\quad
{\bf S_2} = (0,1,0,0),\quad
{\bf S_3} = (0,0,1,0),\quad
{\bf S_4} = (0,0,0,1)
\ee
(cf. (2.25)-(2.28)),
one has
\be
({\bf A,B,C,D})
= \frac{1}{4!}(a_1b_2c_3d_4+a_1b_2c_4d_3+\dots+a_4b_3c_2d_1).
\ee
\ses

\nin
{\bf DEFINITION}. The $\cA\cH_4$-image
$(A,B,C,D)$
of
the quadralinear form
({\bf A,B,C,D})
is called
{\it the scalar quadraproduct} in
the $\cA\cH_4$-algebra.
\ses

\nin
The form (3.5) shows the symmetry
$$
({\bf A,B,C,D})
=
({\bf B,A,C,D})
=
({\bf C,B,A,D})
=
\dots
({\bf A,B,D,C})
$$
and the property of linearity:
$$
(\mu {\bf A}+\nu{\bf  E,B,C,D})
=
\mu({\bf A,B,C,D})
+
\nu({\bf E,B,C,D}),\,\dots
$$
(with respect to each argument), where
$\mu,\nu\in\set R$.

Using the initial quadralinear form (3.5),
it is possible to construct in
$\cV\cH_4$
three mixed metric forms from two single
vectors,
proposing
\be
({\bf A, A, A, B})
=
\frac{1}{4}\lf(a_1a_2a_3b_4+a_1a_2a_4b_3
+a_1a_3a_4b_2+a_2a_3a_4b_1\rg),
\ee
\ses
\be
({\bf A, A, B, B}) = \frac{1}{6}\lf(a_1a_2b_3b_4
+
a_1a_3b_2b_4+a_1a_4b_2b_3
+
a_2a_3b_1b_4+a_2a_4b_1b_3
+
a_3a_4b_1b_2\rg),
\ee
and
\ses
\be
({\bf A, B, B, B})
=
\frac{1}{4}\lf(a_1b_2b_3b_4+a_2b_1b_3b_4+a_3b_1b_2b_4+
a_4b_1b_2b_3\rg).
\ee
\ses

In case of {\it unit vectors},
\be
{\bf a}=\frac{{\bf A}}{|{\bf A}|},\quad
{\bf b}=\frac{{\bf B}}{|{\bf B}|},\quad
{\bf c}=\frac{{\bf C}}{|{\bf C}|}, \quad
{\bf d}=\frac{{\bf D}}{|{\bf D}|},
\ee
so that
\be
({\bf a,a,a,a})=
({\bf b,b,b,b})=
({\bf c,c,c,c})=
({\bf d,d,d,d})=1,
\ee
\ses
we get
\be
({\bf a, a, a, b}) = \frac{({\bf A, A, A, B})}{|{\bf A}|^3|{\bf B}|},
\ee
\ses
\be
({\bf a, a, b, b}) = \frac{({\bf A, A, B, B})}{|{\bf A} |^2|{\bf B} |^2},
\ee
\ses
\be
({\bf a, b, b, b}) = \frac{({\bf A, B, B, B})}{|{\bf A} ||{\bf B} |^3},
\ee
\ses
\be
({\bf a, b, c, d}) =
\frac{({\bf A, B, C, D})}{|{\bf A} ||{\bf B} ||{\bf C} ||{\bf D} |},
\ee
and
\ses
\be
({\bf a, a, a, b})
=
\frac{1}{4}\lf(\frac{a_1a_2a_3}{b_1b_2b_3}+\frac{a_1a_2a_4}{b_1b_2b_4}
+\frac{a_1a_3a_4}{b_1b_3b_4}+\frac{a_2a_3a_4}{b_2b_3b_4}\rg),
\ee
\ses
\ses
\ses
\be
({\bf a, a, b, b}) = \frac{1}{6}\lf(\frac{a_1a_2}{b_1b_2}+
\frac{a_1a_3}{b_1b_3}+\frac{a_1a_4}{b_1b_4}+
\frac{a_2a_3}{b_2b_3}+\frac{a_2a_4}{b_2b_4}
+\frac{a_3a_4}{b_3b_4}\rg),
\ee
\ses
\ses
\ses
\be
({\bf a, b, b, b})
=
\frac{1}{4}\lf(\frac{a_1}{b_1}+\frac{a_2}{b_2}+\frac{a_3}{b_3}+
\frac{a_4}{b_4}\rg).
\ee

Given
a
vector $\bf C\in\cV\cH_4$,
which is equal to geometric difference of two other vectors $\bf A\in\cV\cH_4$
and $\bf B\in\cV\cH_4$,
we
may use the numerical values of these metric forms
to
get
for the difference measure
\be
||{\bf C}||=
({\bf C, C, C, C})=({\bf A-B, A-B, A-B, A-B})
\ee
the generalized metric representation:
\be
||{\bf C}||
= ({\bf A, A, A, A})-4({\bf A, A, A, B})+
6({\bf A, A, B, B})-4({\bf A, B, B, B})+({\bf B, B, B, B})
\ee
\ses

\nin
Similarly
as the cosine of an angle between two vectors in an Euclidean space
allows to express the length of their geometric difference through the
lengths of addends, we get
$$
({\bf C,C})
=({\bf A-B,A-B})=({\bf A,A})-2({\bf A,B})+
({\bf B,B})
\equiv
|{\bf A}|^2-2|{\bf A}||{\bf B}|({\bf a,b})+|{\bf B}|^2,
$$
that is,
\ses\\
\be
({\bf C,C})
=|{\bf A}|^2-2|{\bf A}||{\bf B}|\cos({\bf A,B})+|{\bf B}|^2.
\ee

\ses

\nin
Therefore,
the concept of hyperbolic functions
can be extended to include
the set of four
{\it quadrahyperbolic cosines}:
\be
\cosh ({\bf A, A, A, B}),\quad
 \cosh ({\bf A, A, B, B}),\quad
 \cosh ({\bf A, B, B, B}),\quad
\cosh ({\bf A, B, C, D})
\ee
meaningful in the
space $\cV\cH_4$.
These trigonometric functions
are scalar functions
of four unit vectors,
or three real arguments, in contrast to
the bilinear theories in which the trigonometric functions depend on
but two vectors.

Finally, the pseudo-Euclidean concept of the pseudosphere
(of the hyperboloid)
can be extended according to
\ses

\nin
{\bf DEFINITION}.
The 3-dimensional hypersurface
\be
\cI_4=\{{\bf A}\in\cV\cH_4:\,||{\bf A}||=1, \, a_1>0, a_2>0, a_3>0, a_4>0\}
\ee
is called {\it the $\cH_4$-hyperboloid}.
\ses

\nin
Generally, due analogs of the above definition should be
applied to each of 16 sectors of the space
$\cV\cH_4$;
we, however, restrict our treatment to the $a_1'$-oriented sector
for definiteness, unless otherwise stated explicitly.

Extension of the pseudosphere of radius $r$ should read
\be
\cI_4(r)=
\{{\bf A}\in\cV\cH_4:\, ||{\bf A}||=r>0\,, a_1>0, a_2>0, a_3>0, a_4>0\}.
\ee
This hypersurface may naturally be called
{\it the $\cH_4$-hyperboloid of radius $r$}.
In accord with this terminology we may refer to (3.22)
as to
{\it the unit $\cH_4$-hyperboloid}:
\be
\cI_4=\cI_4(1).
\ee
\ses
\ses\\

\setcounter{sctn}{4}
\setcounter{equation}{0}

{\nin\bf 4. Transversality, Orthogonality, and
Angle
in  ${\bf\cV\cH_4}$}
\ses

Let us introduce
\ses

\nin
{\bf DEFINITION}.
A vector
${\bf B}$
is said {\it transversal} to a vector
${\bf A}$
if the
metric
form
$({\bf A, A, A, B})$ is equal to zero.
If additionally
the
alternative
metric form
$({\bf A,B,B,B})$ is equal
to zero, then the vectors ${\bf A}$ and ${\bf B}$
are called {\it mutually transversal}.
\ses

\nin
The
nature of a quadrahyperbalic space proves to be such
that there are no mutually
transversal unit vectors for which the metric form
$({\bf A, A, B, B})$
would also
vanish.
Instead they may equal its
extreme
values
$+\frac{1}{3}$ or $-\frac{1}{3}$.

Let's adopt to call
two nonisotropic
vectors of a quadrahyperbolic space {\it orthogonal},
if the metric forms
derivated therefrom
obey
the conditions
\be
({\bf A,A,A,B})=0,\quad
({\bf A,A,B,B})=-\fr{1}{3}
|{\bf A}|^2
|{\bf B}|^2,
\quad
({\bf A,B,B,B})=0.
\ee
In view of (3.18),
this entails
for two orthogonal vectors of quadrahyperbolic
space
the generalized representation
\be
|{\bf C}|^4
=
|{\bf A}|^4-2|{\bf A}|^2|{\bf B}|^2+|{\bf B}|^4=(|{\bf A}|^2-|{\bf B}|^2)^2,
\ee
so that an analog of the Pythagorean theorem reads as
\be
|{\bf C}|^2=|{\bf A}|^2-|{\bf B}|^2
\ee
(here, the vectors ${\bf A}$
and ${\bf B}$ may belong to various sectors of the space
$\cV\cH_4$),
which outwardly
coincides with the expression for orthogonal vectors in a
pseudo-Euclidean
space.

If four noncoplanar vectors of a quadrahyperbolic space satisfy
pairwisely
the
properties (4.1) and, additionally,
have unit lengths, we say that the
vectors comprise
{an orthonormal basis}
(which is a counterpart of
an orthonormal basis in a bilinear space).

Whether the basic units  $1$, $I$, $J$, $K$, when changed for their bold
counterparts
${\bf 1}$, ${\bf I}$, ${\bf J}$, ${\bf K}$,
form
an orthonormal basis  of a
quadrahyperbolic space under study?
Since direct calculations yield
the values
\be
({\bf 1,1,1,1}) = 1, \quad
({\bf I, I, I, I}) = 1,\quad
 ({\bf J, J, J, J}) = 1,\quad
 ({\bf K, K, K, K}) = 1,
\ee
\ses
\be
({\bf 1,1,1,I}) = ({\bf 1,1,1, J}) = \dots  = ({\bf 1, K, K, K}) = 0,
\ee
\ses
\be
({\bf  1,1, I, I}) = ({\bf 1,1, J, J}) =\dots = ({\bf J, J, K, K})
=
-\frac{1}{3},
\ee
\ses
\be
({\bf  1, I, J, J}) = ({\bf 1, I, K, K}) =\dots = ({\bf I, J, K, K}) = 0,
\ee
\ses
\be
({\bf  1, I, J, K}) = \frac{1}{3},
\ee
the answer to the above question should be said in positive.

Any other quadruples of vectors giving rise to the same
values for metric forms
may be
summarizingly expressed by the following conditions:
\be
({\bf E_p,E_q,E_r,E_s})=
\left\{
\begin{array}{cccr}
1 & if & p=q=r=s,\\
0 & if  & (p=q=r)\ne s,\\
-\frac{1}{3} & if & (p=q)\ne(r=s),\\
0 & if & p=(q\ne r \ne s),\\
\frac{1}{3} & if & (p\ne q\ne r\ne s).\\
\end{array}
\right.
\ee
\ses

\nin
They will also make an orthonormal basis; the parentheses
in the right-hand
side mean
all possible
permutations.

With respect to the absolute basis ${\bf S_1, S_2, S_3, S_4}$,
components of each vector ${\bf A}\in\cV\cH_4$ are found to be
$
24({\bf A, S_2, S_3, S_4}),\,
24({\bf S_1, A, S_3, S_4}),\,
24({\bf S_1, S_2, A, S_4}),\,
24({\bf S_1, S_2, S_3, A}).
$
Whence we arrive
at
the expansion
\be
A =24\Bigl(({\bf A,S_2,S_3,S_4}){\bf S_1}+
({\bf S_1,A,S_3,S_4}){\bf S_2}+
({\bf S_1,S_2,A,S_4}){\bf S_3}+
({\bf S_1,S_2,S_3,A}){\bf S_4}\Bigr).
\ee
If we go over to another basis ${\bf E_q}$,
connected with an absolute one by means of the
transformation
${\bf S_p} = M_{pq}{\bf E_q}$,
the expansion of the vector ${\bf A}$ will modify to read
\be
{\bf A} = 24 (({\bf A}, M_{2q}{\bf E_q}, M_{3q}{\bf E_q},
M_{4q}{\bf E_q}) M_{1q}{\bf E_q} + (M_{1q}{\bf E_q}, {\bf A},
M_{3q}{\bf E_q}, M_{4q}{\bf E_q}) M_{12q}{\bf E_q}
\ee
\ses
\be
 + (M_{1q}{\bf E_q}, M_{2q}{\bf E_q}, {\bf A}, M_{4q}{\bf E_q}) M_{3q}{\bf E_q}
 + (M_{1q}{\bf E_q}, M_{2q}{\bf E_q}, M_{3q}{\bf E_q, A})M_{4q}{\bf E_q}).
\ee

In particular, for the orthonormalized
basis ${\bf 1}$, ${\bf I}$, ${\bf J}$, ${\bf K}$ the remarkable representation
\be
{\bf A} = ({\bf A, 1,1,1}){\bf 1} + ({\bf A, I, I, I}){\bf  I} +
({\bf A, J, J, J}){\bf J} + ({\bf A, K, K, K}){\bf K}
\ee
can be obtained after simple direct calculations.
Thus, the forms $({\bf A, 1,1,1})$, $({\bf A, I, I, I})$,
$({\bf A, J, J, J})$, and
$({\bf A, K, K, K})$
assign
the connection of vector ${\bf A}$
components in an orthonormalized basis
with
the product $({\bf A, B, C, D})$,
and therefore these forms can
be interpreted
as
{\it
transversal
projections}
 of the given vector on the
axes
${\bf 1}$, ${\bf I}$,
${\bf J}$, ${\bf K}$.

Four respective
transversal projections of a unit vector ${\bf a}\in\cV\cH_4$ on an
orthonormal basis's axes
prove to be
connected by the following relation
$$
\cosh^4 ({\bf a, 1,1,1}) + \cosh^4 ({\bf a, I, I, I}) +
\cosh^4 ({\bf a, J, J, J}) +
\cosh^4 ({\bf a, K, K, K})
$$
\ses
$$
 -
2\cosh^2 ({\bf a, 1,1,1}) \cosh^2 ({\bf a, I, I, I})
-2\cosh^2 ({\bf a, 1,1,1}) \cosh^2 ({\bf a, J, J, J})
$$
\ses
$$
 -2\cosh^2 ({\bf a, 1,1,1})
\cosh^2({\bf a, K, K, K})
-
2\cosh^2({\bf a, I, I, I})\cosh^2 ({\bf a, J, J, J})
$$
\ses
$$
-2\cosh^2({\bf a, I, I, I})
\cosh^2({\bf a, K, K, K})
-2\cosh^2({\bf a, J, J, J})\cosh^2({\bf a, K, K, K})
$$
\ses
\be
+8\cosh ({\bf a, 1,1,1})\cosh({\bf a, I, I, I})\cosh({\bf a, J, J, J})
\cosh ({\bf a, K, K, K})=1,
\ee
which, as a matter of fact, is
an analog of Pythagorean theorem and extend
the latter
for a unit
diagonal of a quadrahyperbolic parallelotope which edges are pairwisely
orthogonal.

By an
{\it $\cH_4$-angle}
$\phi$
between two vectors
${\bf A}$ and ${\bf B}$ we shall naturally mean
the length
of a geodesic arc which joins over
the unit $\cH_4$-hyperboloid
$\cI_4$
the points of intersection of the vectors (or their straight continuations)
with
$\cI_4$.
It is possible to show that its numerical
value is determined
by
\be
\phi\eqdef\Bigl|\ln({\bf A})-\ln({\bf B})\Bigr|
=
\Bigl|({\alpha}_A-{\alpha}_B){\bf I}+
({\beta}_A-{\beta}_B){\bf J}+
({\gamma}_A-{\gamma}_B){\bf K}\Bigr|,
\ee
where $\ln({\bf A})$ means the
logarithmic function (2.42) of the quadrahyperbolic
variable $A$ that relates to the vector ${\bf A}$.
We remain it to the reader to verify the validity of the
component
representation
\be
\phi=\sqrt[4]{
\Bigl|\ln\frac{a_1}{b_1}
\ln\frac{a_2}{b_2}\ln\frac{a_3}{b_3}\ln\frac{a_4}{b_4}\Bigr|}\,,
\ee
which implies the relations
$a_1a_2a_3a_4=1$
and
$b_1b_2b_3b_4=1$.
\ses
\ses\\

\setcounter{sctn}{5}
\setcounter{equation}{0}

{\nin\bf 5. Arcs, ${\bf\cV\cH_4}$-{\bf Sectors,}
${\bf\cV\cH_4}$-{\bf Cones, and
Isometries}
\rm
\ses

We
introduce
\ses

\nin
{\bf DEFINITION}. Given two points on the $\cH_4$-hyperboloid (3.22),
the
geodesic piece
that joins the points is called an
{\it  arc}.
\ses

\nin
{\bf DEFINITION}.
A two-dimensional surface formed by two
vectors
$\bf A$ and $\bf B$
subject to the condition that the ends of the vectors
are connected by an arc
will be called
{\it the
$\cV\cH_4$-sector},
to be denoted as
$\cT_4({\bf A}, {\bf B})$.
If the vectors are unit, the adjective ${\it unit}$
will be added to the names of these sectors.
\ses

\nin
This expands the corresponding concept used in bilinear spaces,
where triangles and sectors
are always plane figures;
a $\cV\cH_4$-sector is generally not a plane figure,
-
being rather a ``cone-type surface".
\ses

\nin
{\bf DEFINITION}. The
$\cV\cH_4$-{\it cone}
$\cC_4(r)$
is a two-dimensional surface which generatrix is a semiline issued from the
origin of the space
$\cV\cH_4$
and which intersection with the unit $\cH_4$-hyperboloid
$\cI_4$
is an $r$-radius circle drawn on
the
$\cI_4$.
\ses

\nin
The latter definition provides an extension of
the ordinary Euclidean
cone of rotation.
\ses

\nin
{\bf DEFINITION}. Two
figures formed by a set of vectors of equal number are called {\it isometric}
if all respective scalar products, in the sense of the definitions
introduced by the list (3.6)-(3.8), have equal values.
\ses

\nin
In accordance with the
latter definition,
given two pairs of vectors
of a quadralinear space,
${\bf A}$ and ${\bf B}$,
{\it resp.}
${\bf A'}$ and ${\bf B'}$,
the figure made up by the first two vectors will be isometric
to the figure made up by the last two figures, when
all the equalities
\be
({\bf A,A,A,A})
=
({\bf A',A',A',A'}),
\ee
\ses
\be
({\bf A,A,A,B})
=
({\bf A',A',A',B'}),
\ee
\ses
\be
({\bf A,A,B,B})
=
({\bf A',A',B',B'}),
\ee
\ses
\be
({\bf A,B,B,B})
=
({\bf A',B',B',B'}),
\ee
and
\be
({\bf B,B,B,B})
=
({\bf B',B',B',B'})
\ee
hold simultaneously.
In particular,
two vectors
${\bf A}$ and ${\bf A'}$
of a quadralinear space
are isometric if
\be
({\bf A,A,A,A})
=
({\bf A',A',A',A'})
\ne 0.
\ee
\ses
\ses\\

\setcounter{sctn}{6}
\setcounter{equation}{0}

{\nin\bf 6. $\cH_4$-Holomorphic Functions}
\ses

Given {\it a function of quadrahyperbolic variable}:
\be
f(A)=U
(a_1', a_2', a_3', a_4')
+
V(a_1', a_2', a_3', a_4')I
+
W( a_1', a_2', a_3', a_4')J
+Q
( a_1', a_2', a_3', a_4')K,
\ee
\ses
where
$U,V,W,Q$
are smooth functions of four
real arguments,
we treat the
set
$U,V,W,Q$
 naturally as {\it the hypercomplex components} of the function $f$.
\ses

\nin
{\bf DEFINITION}. The function (6.1) is called
{\it $\cH_4$-holomorphic}, if the following
{\it $\cH_4$-holomorphic conditions}
\be
\D U{a_1'}=
\D V{ a_2'}=
\D W{ a_3'}=
\D Q{ a_4'},
\qquad
\D U{ a_2'}=
\D V{ a_1'}=
\D W{ a_4'}=
\D Q{ a_3'},
\ee
\ses
\ses
\ses
\ses
\be
\D U{a_3'}=
\D V{ a_4'}=
\D W{a_1'}=
\D Q{ a_2'},
\qquad
\D U{ a_4'}=
\D V{ a_3'}=
\D W{ a_2'}=
\D Q{a_1'}
\ee
hold fine.

To elucidate the meaning of the latter conditions, let us
consider the direct differentials
\be
\De A
=
\De a_1'1
+
\De a_2'I
+
\De a_3'J
+
\De a_4'K
\ee
and
\ses
\be
\De f=\De U+\De VI
+\De WJ+\De QK,
\ee
and introduce the partial derivative
\be
\prtl f=\{\prtl_a f\}:
\ee
\ses
\ses
\be
\prtl_1 f=
\D U{a_1'}
+
\D V{a_1'}I
+
\D W{a_1'}J
+
\D Q{a_1'}K,
\ee
\ses
\ses
\ses
\be
\prtl_2 f=
\D U{a_2'}
+
\D V{a_2'}I
+
\D W{a_2'}J
+
\D Q{a_2'}K,
\ee
\ses
\ses
\ses
\be
\prtl_3 f=
\D U{a_3'}
+
\D V{a_3'}I
+
\D W{a_3'}J
+
\D Q{a_3'}K,
\ee
\ses
\ses
\ses
\be
\prtl_4 f=
\D U{a_4'}
+
\D V{a_4'}I
+
\D W{a_4'}J
+
\D Q{a_4'}K.
\ee
\ses

\nin
Now we want to retain the ordinary multiplication rule
\be
\De f=\De A\cdot\prtl f.
\ee
It can readily be verified that
{\it
Eqs.}
(6.4)-(6.11)
{\it
entail the $\cH_4$-conditions}
(6.2)-(6.3).
\ses
\ses\\

\setcounter{sctn}{7}
\setcounter{equation}{0}

{\nin\bf 7. Relationship with Finsler Geometry}
\ses

We may identify the Finslerian metric function
\be
F({\bf A})=\sqrt[4]{|({\bf A,A,A,A})|}
\ee
with the length (3.3).
Owing to (2.40), we can use the component representation
\be
F({\bf A})=\sqrt[4]{|a_1a_2a_3a_4|}
\ee
which is identical with the
known
representation
of the
{\it Berwald-Moor}
metric function (see [25]).
If we construct the associated covariant components
\be
y_p\eqdef
\fr12\D{F^2}{a^p},
\ee
where
$a^p=\{a_1,a_2,a_3,a_4\}$,
we get
\be
y_p=\fr{F^2}{4a^p}.
\ee
Farther calculation of the Finslerian metric tensor
\be
g_{pq}\eqdef
\fr12
\Dd{F^2}{a^p}{a^q}
\ee
yields
\be
g_{pq}
=
\fr{2y_py_q}{F^2}-\fr{F^2}{4a^pa^q}\de_{pq},
\ee
where
$\de$
stands for the Kronecker symbol.

By comparing (2.24) with (7.4)
we may conclude that the Finslerian concept of covariant components
is tantamount to the concept
of the inverse number in the quadralinear space.

Also, {\it
the Finslerian indicatrix} given by
\be
F({\bf A})=1
\ee
is equivalent to the $\cH_4$-hyperboloid (3.22).
Many other analogies can be traced farther.
\ses
\ses\\

\setcounter{sctn}{8}
\setcounter{equation}{0}

{\nin\bf 8. Conclusions
and
Prospects}
\ses

Various significant cases of polyspaces can be specified to
make interesting applications.
Under quadralinear symmetric-type treatment,
the fundamental metric forms
\be
||A||={a_1}^4+{a_2}^4+{a_3}^4+{a_4}^4,
\ee
\ses
\be
||A||=(a_2+a_3+a_4){a_1}^3+(a_1+a_3+a_4){a_2}^3+
(a_1+a_2+a_4){a_3}^3+(a_1+a_2+a_3){a_4}^3,
\ee
\ses
\be
||A||={a_1}^2{a_2}^2+
{a_1}^2{a_3}^2+{a_1}^2{a_4}^2+{a_2}^2{a_3}^2+{a_2}^2{a_4}^2+{a_3}^2{a_4}^2,
\ee
\ses
$$
||A||=(a_2a_3+a_2a_4+a_3a_4){a_1}^2+(a_1a_3+a_1a_4+a_3a_4){a_2}^2
$$
\ses
\be
+
(a_1a_2+a_1a_4+a_2a_4){a_3}^2+(a_1a_2+a_1a_3+a_2a_3){a_4}^2,
\ee
\ses
\be
||A||=a_1a_2a_3a_4
\ee
seem to be most attractive
because of their simple algebraic structures.
In the present work,
we have opened up due possibilities inhereted in the last choice (8.5).
Obviously, the above forms can be considered as ``canonical
ingredients"
for farther classification of quadralinear spaces.
The forms (8.1)-(8.5)
are
algebraically
independent,
-
in the sense that no member of the list (8.1)-(8.5)
can be mapped under a linear
transformation in another member of this list.
Therefore, the forms may
serve as
canonical elements for
classification
of
quadralinear spaces;
extension to any two-, three-, or $(n>4)$-dimensional theories
is straightforward in many aspects.

Because of the apparent mathematical
simplicity and beauty, and also rather nontrivial geometric
structure of the polyspaces considered above,
it can be hoped
that the content and subject of the present paper
may be of interest also to those who develop and study new
physical generalized-relativisitic approaches rather than proper algebra of
polynumbers.
\ses
\ses\\

{\nin\bf  Research Problems}
\ses

\nin
There is much interesting work still to be done on the ideas
and methods
proposed above.
The following list of nearest pending problems
may be set forth to resolve, -  which
way, whenever being successful, would
favour various practical as well as
relativistic scopes of applications
of
commutative and associative hypercomplex numbers and multilinear spaces
treated above.
\ses

\nin
1. Develop an appropriate systematization
for polynumbers
of square-type nature.

Classify polynumbers which fail to be of square-type nature.
\ses

\nin
2. Offer a simple algorithm of constructing mutual-conjugates to match a
maximally

broad set of polynumbers.
\ses

\nin
3. Propose
self-consistent
geometric treatment of arguments
$\al, \beta, \ga$
(see Eqs. (2.44)-

(2.49))
of exponential
expression of
quadrahyperbolic
(and other)
polynumbers.

Do the arguments
admit a meaning of geometrical angles proper?\ses

\nin
4. Find approporiate
extended
$\cH_4$-rotation transformations. Can the
sectors $\cT_4$
or the

cones $\cC_4$
in the
$\cV\cH_4$-space
be moved subject to the condition
that the lengths

of generating
vectors
and
arcs remain unchanged?
\ses

\nin
5.Trace the possibilities to have congruent transformations in the
spaces
under

study.
Is it possible to find the congruent transformations such that isometric

figures
(see the conditions
(5.1)-(5.5))
can be moved to identify?
\ses

\nin
6. In multilinear
spaces, define additive metric parameters for figures made up of

three and more vectors.
\ses

\nin
7. Propose and develop, up to isomorphism, classification of four-dimensional
spaces

with quadralinear
symmetrical forms.
\ses

\nin
8. For the
spaces under study, investigate existence of two-dimensional hyperplanes,

as well as three-dimensional hypersurfaces, such that they are endowed
with

a
Riemannian-type metric.
\ses

\nin
9. Ellucidate the structure of holomorphic functions of polynumbers and develop

respective geometric treatments.
Study
singular points, lines, and regions

appeared under
analitic mappings of
associated hypercomplex manifolds.
\ses

\nin
10.Describe due extensions for arbitrary dimensions.
\ses

\nin
11.Seek to develop novel generalized physical aspects, in particular those
referred

to light behaviour, basig on the $\cA\cH_4$-algebra
or on the
$\cV\cH_4$-space.
\ses

\nin
12. Understand and investigate
connection of Pythagorean theorem analogs for length

of diagonals of parallelotopes in multilinear spaces with Diophantine
equations.
\ses
\ses\\

\def\bibit[#1]#2\par{\rm\noindent\parskip1pt
                     \parbox[t]{.05\textwidth}{\mbox{}\hfill[#1]}\hfill
                     \parbox[t]{.925\textwidth}{\baselineskip11pt#2}\par}

\nin{\bf References}
\bigskip

\bibit[1] H. Helmholtz:\it Ueber die Thatsachen  die der Geometrie zum
Grunde liegen, \rm G\"{o}ttingen, Nachr. 1868.

\bibit[2] I.L. Kantor, A.S. Solodovnikov: {\it
Hypercomplex numbers: An
elementary introduction to algebras}, Springer, Berlin 1989.

\bibit[3] G. Birkhoff and S. MacLane: \it
Modern Algebra, \rm
Macmillan, New York, Third Edition
1965.

\bibit[4] B.L. van der Waerden:
\it Modern Algebra, \rm F.Ungar, New York, Third Edition 1950.

\bibit[5] O. Taussky:
\it
Algebra, \rm
\ in Handbook of Physics, ed. by E. U.Condon and H. Odishaw,
McGraw-Hill, New York, Second Edition 1958.

\bibit[6] J.W. Brown and R.V.Churchill: \it Complex variables and
applications, \rm McGraw-Hill,  New York, 1996.

\bibit[7] R.L. Goodstein: \it Complex functions, \rm McGraw-Hill,  New York,
1965.

\bibit[8] K. G\"{u}rlebeck and W.Spr\"{o}ssig: \it Quaternionic and
Clifford calculus for physicists and engineers, \rm John Wiley\&Sons, 1997.

\bibit[9] G.B. Price:
\it
An introduction to multicomplex spaces and
functons, \rm
Marcel Dekker, 1991.

\bibit[10] J.B. Seaborn: \it
Hypergeometric functions and their applications,
\rm
Springer, Berlin 1991.

\bibit[11] A. van Proeyen:
Special geometries, from real to quaternionic,
hep-th/0110263.

\bibit[12] Y. Nutku, M.B. Sheftel:
A family of heavenly metrics,
gr-qc/0105088.

\bibit[13] G. Papadopoulos:  Brane Solitons and Hypercomplex Structures,
 math.DG/0003024.

\bibit[14] D.M.J. Calderbank
and P. Tod:
Einstein metrics, hypercomplex structures and the Toda field equation,
math.DG/9911121.

\bibit[15] M. Ansorg:
Differentially rotating disks of dust: Arbitrary rotation law,
 gr-qc/0006045.

 \bibit[16] S. Ollariu:
Complex numbers in n dimension,
 math.CV/9011077.

\bibit[17] D.M. J. Calderbank,
Einstein metrics, hypercomplex structures and the Toda field equation,
P. Tod, math.DG/9911121.

\bibit[18] D.L. Stefano:
Hypercomplex group theory,
 physics/9703033.

\bibit[19] A. Dimakis and
 F. M\"{u}ller-Hoissen:
Bicomplex and B\"{a}cklund transformations,
 nlin.SI/0104071.

\bibit[20] D. Joyce:
A theory of quaternionic algebra, with applications to
hypercomplex geometry,
 math.DG/0010079.

\bibit[21] S. R\"{o}nn:
Bicomplex algebra and function theory,
 math.CV/0101200.

\bibit[22] D.L. Stefano and R. Pietro:
Local Hypercomplex Analyticity,
funct-an/9703002.

\bibit[23] A. Sudbery:
Quaternionic analysis,
\it Math. Proc. Camb. Phil. Soc. \rm
\bf 85 \rm (1979),
199.

\bibit[24] H. Rund: \it The Differential Geometry of Finsler
 Spaces, \rm Springer, Berlin 1959.

\bibit[25] G.S. Asanov: \it Finsler Geometry, Relativity and Gauge
 Theories, \rm D.~Reidel Publ. Comp., Dordrecht 1985.

\bibit[26] G.S. Asanov:
 Finsleroids reflect future-past asymmetry.
\it Rep. Math. Phys.
\bf 47 \rm (2001), 323.

\bibit[27] G.S. Asanov: Can neutrinos and high-energy particles test
Finsler metric of space-time?
hep-ph/0009305.

\bibit[28] G.S. Asanov:
Finslerian extension of Lorentz transformations
and first-order censorship theorem.
\it Found. Phys. Lett. \bf 15 \rm (2002), 199.

\bibit[29] G.S. Asanov:
Finslerian anisotropic relativistic metric function obtainable
under breakdown of rotational symmetry.
gr-qc/0204070.

\end{document}